\documentclass[acmsmall]{acmart}
\usepackage{amsmath}
\usepackage{mathtools, cuted}
\usepackage{lipsum, color}
\usepackage{breqn}
\usepackage{rotating}
\usepackage{pdflscape}
\usepackage{subcaption}

\AtBeginDocument{%
  \providecommand\BibTeX{{%
    \normalfont B\kern-0.5em{\scshape i\kern-0.25em b}\kern-0.8em\TeX}}}

\setcopyright{acmcopyright}
\copyrightyear{2020}
\acmYear{2020}
\acmDOI{10.1145/1122445.1122456}

\acmJournal{JACM}
\acmVolume{37}
\acmNumber{4}
\acmArticle{111}
\acmMonth{1}

\begin{document}

\title{Origins of Algorithmic Instabilities in Crowdsourced Ranking}

\author{Keith Burghardt}
\affiliation{%
  \institution{USC Information Sciences Institute, USA}
  \streetaddress{4676 Admiralty Way}
  \city{Marina del Rey}
  \state{Caliornia}
  \postcode{90292}}
\email{keithab@isi.edu}
\author{Tad Hogg~}
\affiliation{%
  \institution{Institute for Molecular Manufacturing, USA}
  \streetaddress{555 Bryant Street, Suite 354}
  \city{Palo Alto}
  \state{Caliornia}
  \postcode{94301}}
\email{tadhogg@yahoo.com}
\author{Raissa M. D'Souza}
\affiliation{
 \institution{University of California, Davis, USA}
\streetaddress{1 Shields Ave}
  \city{Davis}
  \state{Caliornia}
  \postcode{95616}}
\affiliation{
 \institution{Santa Fe Institute, USA}
\streetaddress{1399 Hyde Park Road}
  \city{Santa Fe}
  \state{New Mexico}
  \postcode{87501}}
\email{raissa@cse.ucdavis.edu}
\author{Kristina Lerman}
\affiliation{%
  \institution{USC Information Sciences Institute, USA}
 }
\email{lerman@isi.edu}
\author{M\'arton P\'osfai}
\affiliation{%
  \institution{Central European University, Hungary}
  \streetaddress{N\'ador u. 9}
  \city{Budapest}
  \state{Hungary}
  \postcode{1051}}
\affiliation{
 \institution{University of California, Davis, USA}
}
  \email{posfaim@ceu.edu}
\renewcommand{\shortauthors}{Keith Burghardt et al.}

%
\begin{abstract}
Crowdsourcing systems aggregate decisions of many people to help users quickly identify high-quality options, such as the best answers to questions or interesting news stories. A long-standing issue in crowdsourcing is how option quality and human judgement heuristics interact to affect collective outcomes, such as the perceived popularity of options. We address this limitation by conducting a controlled experiment where subjects choose between two ranked options whose quality can be independently varied. We use this data to construct a model that quantifies how judgement heuristics and option quality combine when deciding between two options. The model reveals popularity-ranking can be unstable: unless the quality difference between the two options is sufficiently high, the higher quality option is not guaranteed to be eventually ranked on top. To rectify this instability, we create an algorithm that accounts for judgement heuristics to infer the best option and rank it first. This algorithm is guaranteed to be optimal if data matches the model. When the data does not match the model, however, simulations show that in practice this algorithm performs better or at least as well as popularity-based and recency-based ranking for any two-choice question. Our work suggests that algorithms relying on inference of mathematical models of user behavior can substantially improve outcomes in crowdsourcing systems.
\end{abstract}

%
%


%

%

\setcopyright{acmlicensed}
\acmJournal{PACMHCI}
\acmYear{2020} \acmVolume{4} \acmNumber{CSCW2} \acmArticle{166} \acmMonth{10} \acmPrice{15.00}\acmDOI{10.1145/3415237}
%
\maketitle

\section{Introduction}
Crowdsourcing websites aggregate judgments in order to 
help users discover high quality content. These systems typically combine choices of many people to algorithmically \textit{rank} content so that better items---product reviews~\cite{lim2015evaluating}, news stories~\cite{PopularDynam,SocialInfluenceBias}, or answers on question-answering (Q\&A) platforms~\cite{BestAnswerYahoo,HighQualityQA}---are easier to find.

Despite a long history of crowdsourcing~\cite{JuryThm,Galton1908}, some of its limitations have only recently become apparent. Salganik et al. (\citeyear{Salganik2006}) found that aggregating the votes of many people to rank songs increases the inequality and instability of song popularity. Even when starting from the same initial conditions, the same songs could end up with vastly different rankings. Other studies have shown that algorithmic ranking can amplify the inequality of popularity~\cite{lerman14as,Burghardt2018} and bias collective outcomes in crowdsourcing applications~\cite{MyopiaCrowd,dev2019quantifying}. In addition, information about the choices of other users affects decisions in complex ways~\cite{SocialInfluenceBias,Hogg2015hcomp,talton2019people}. Unfortunately for crowdsourcing system designers, it is still not clear how these finding could help improve collective outcomes, in large part due to difficulty of quantifying the quality of options (e.g., the best answer to a question) and its impact on individual decisions. 

To better understand and improve collective outcomes that emerge from individual decisions, we break down the crowdsourcing task into its basic elements: item quality, item ranking, and social influence. We create a controlled experiment to study how these elements jointly affect individual decisions and collective outcomes. We use experimental data to construct and validate a mathematical model of human judgements, and then use it to explore algorithmic ranking and identify strategies to improve crowdsourcing performance. Our study addresses the following research questions:
\begin{description}
    \item[RQ1] How does quality and presentation of options jointly impact individual decisions?
    \item[RQ2] When is algorithmic ranking unstable and does not reliably identify the best option? 
    \item[RQ3] How can we stabilize algorithmic ranking such that the best option is typically ranked first?
\end{description}

\begin{figure}[tbh!]
\centering
\includegraphics[width=0.8\columnwidth]{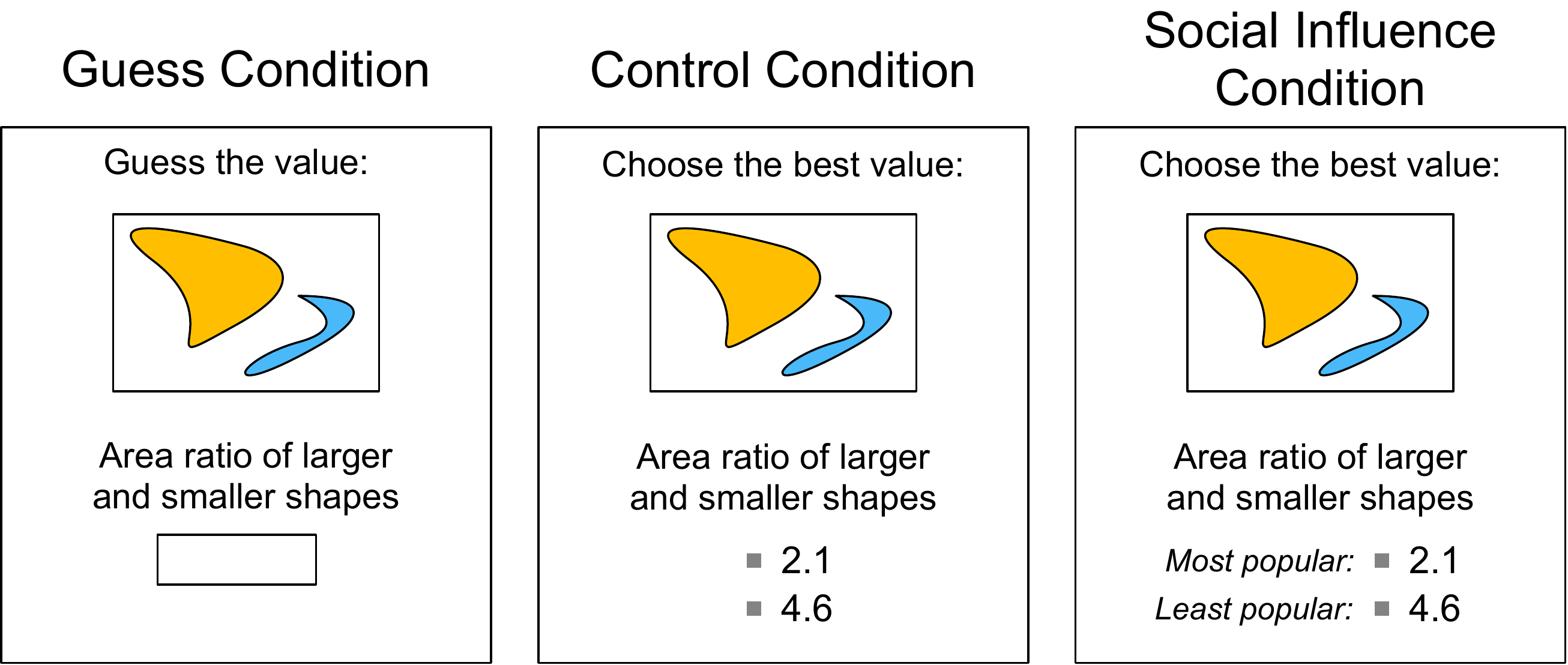}
\caption{\label{fig:ExperimentSchematics} Schematic of the experiment conditions. Left: guess condition, where subjects write a guess for the correct value. Center: control condition, where subjects choose the best of two answers. Right: social influence condition, where a randomly chosen answer is called the ``most popular'' and is ranked first.
}
\end{figure}

Our experiment asks subjects to choose the best answer to questions with objectively correct answers, such as the number of dots in a picture. Figure~\ref{fig:ExperimentSchematics} illustrates one such question, where we ask users to find the area ratio between the largest and smallest shapes. (Other questions used in the experiment are shown in Appendix Fig.~\ref{fig:Questions}.) While these simple questions abstract away some of the complexity of the often-subjective decision making people do in crowdsourcing systems, they allow quality to be objectively measured and its effects on decisions better understood.

The experiment has three conditions shown in Fig.~\ref{fig:ExperimentSchematics}. In the first condition, we let subjects write their answers to understand how their subjective guesses deviate from the correct answers. In the remaining conditions, we ask subjects to choose the best of two randomly generated answers that are randomly ordered. In the control condition, subjects are not told how they are ordered, while in the social influence condition, the first answer is labeled ``more popular''. These simple conditions allow us to disentangle the elements of crowdsourcing systems and begin quantifying how individual decisions affect collective outcomes.

To begin answering RQ1, we construct a mathematical model of the probability to choose an option as a function of its position and quality. The model requires only two parameters to measure \textit{cognitive heuristics} (mental shortcuts people use to make quick and efficient judgements) and is in excellent agreement with experimental data. The first parameter reflects a user's preference to pick the first answer (known as ``position bias''~\cite{lerman14as,Burghardt2018,MusicLabModel,PopularDynam}) and the other parameter measures the rate at which answers are guessed at random. Subjects otherwise pick an answer closest to their initial (unobserved) guess. We call this model the Biased Initial Guess (BIG) model. The BIG model demonstrates that the ``social influence'' experiment condition enhances position bias~\cite{Burghardt2018}, 
therefore cognitive heuristics that produce position bias and social influence can be quantified with a single parameter, a substantial simplification over previous work \cite{MusicLabModel,PopularDynam}. Moreover, it helps explain why users often choose the worst answer when answer quality differences are small.

The BIG model not only improves our understanding of how answers are chosen, but it allows us to test different ranking policies in simulations to answer RQ2 and RQ3. Importantly, these simulations demonstrate that cognitive heuristics can make popularity-based ranking highly unstable.  When the quality difference between the options is small, initially minor differences in popularity can create a cumulative advantage \cite{Rijt2014}, meaning the better answer does not always become the most popular.  However, when the quality difference passes a critical point, popularity-ranking is stable, and the better answer eventually becomes the most popular. These results may help explain an under-appreciated finding of Salganik et al. (\citeyear{Salganik2006}) that the best and worst songs tended to be correctly ranked when songs were ordered by popularity, but intermediate-quality songs landed anywhere in between. 

Finally, to answer RQ3, we propose an algorithm that rectifies this instability by ordering answers based on their inferred quality, which we call RAICR: Rectifying Algorithmic Instabilities in Crowdsourced Ranking. This method is found to be stable and consistently order the better answer first, thus making best answers easier to find, even when they are only slightly better. RAICR ranks answers as well as, or better than, common baselines such as ordering by popularity or by recency (ranking by the last answer picked).

Our work shows that individual decisions within crowdsourcing systems are strongly affected by cognitive heuristics, which collectively create instability and poor crowd wisdom. Designers of crowdsource systems need to account for these biases in order make good content easier to find. Algorithms such as RAICR, however, can correct for these biases, thereby improving the wisdom of crowds. 

\section{Related Literature}

\subsection{Crowdsourcing}
Crowdsourcing has a two-century long history demonstrating how a collective can outperform individual experts \cite{JuryThm,Galton1908,surowiecki2005wisdom,Kaniovski,Simoiu2019}, thus creating the moniker ``wisdom of crowds''. Crowds have been shown to beat sport markets \cite{Brown2019,Peeters2018}, corporate earnings forecasts \cite{Zhi2019}, and improve visual searches \cite{Juni2017}. One reason crowd wisdom works is due to the law of large numbers: assuming unbiased and independent guesses, the average guess should converge to the true value. 

Individual decisions, including in online settings, are biased by cognitive heuristics, such as anchoring \cite{shokouhi2015anchoring}, primacy \cite{Primacy1}, prior beliefs \cite{white2013beliefs} 
and position bias \cite{Burghardt2018}. These biases are not necessarily canceled out with large samples \cite{Prelec2017,Kao2018}. As a result, aggregating guesses of a crowd does not necessarily converge to the correct answer.

Guesses are usually not independent, which can sometimes improve the wisdom of crowds. Social influence models, such as the DeGroot model \cite{Degroot1974}, have been shown to push simulated agents to an optimal decision \cite{Golub2010,Mossel2015,Bala1998,Acemoglu2011}. These results have been backed up experimentally \cite{Becker2017,Becker2019,Ungar2012,Tetlock2017}, even when opinion polarization is included \cite{Becker2019}. One reason social influence can be beneficial is that it encourages people who are way off the mark to improve their guess \cite{Mavrodiev2013,Becker2017,Abeliuk2017www}.

Often, however, social influence can reduce crowd wisdom. Corporate earnings predictions \cite{Zhi2019}, jury decisions  \cite{Kaniovski,BurghardtJury}, and other guesses can degrade with influence \cite{Lorenz2011,Lorenz2015,Simoiu2019}, and malevolent individuals can manipulate people to make particular collective decisions \cite{SocialInfluenceBias,Asch1951}. Too much influence by a single individual can also reduce the wisdom of collective decisions \cite{Becker2017,Acemoglu2011,Golub2010}, and deferring to friends can sometimes make unpopular (and potentially low-quality) ideas appear popular \cite{Lerman2016}. 

Recent work has also demonstrated how other cognitive heuristics can affect crowd wisdom. After the landmark study by Salganik et al. (\citeyear{Salganik2006}), some researchers found that social influence has no effect on decisions \cite{Antenore2018}, or that position bias, i.e., the preference to choose options listed first, largely explain biases in crowdsourcing \cite{lerman14as,MusicLabModel}. Social influence instead enhances the position bias \cite{Burghardt2018,MusicLabModel}. Burghardt et al. (\citeyear{Burghardt2018}) have begun to tease apart these effects, showing that while social influence enhances position bias, it has no marginal effect when we control for position. This is consistent with our approach, which models and rectifies both biases with a single parameter. 

\subsection{Algorithmic Ranking}

The goal of a ranking algorithm is to make good content easier to find. 
Many papers have begun to address this goal \cite{Page1999,Jarvelin2002,Bendersky2011}, which has recently been applied to crowdsourced ranking. Because of human biases, however, algorithms that na{\"i}vely use human feedback to suggest content will end up forming echo chambers \cite{Hilbert2018,Bozdag2013,Hajian2016}, or only recommend already-popular items \cite{Abdollahpouri2017,Abdollahpouri2019}. This can also give some content a cumulative advantage \cite{Rijt2014}, even when it is of similar quality to content that remains unpopular. 

To correct for algorithmic bias in this paper, we create a ranking method that follows the strategy of Watts, who says, ``...we can instead measure directly how they respond to a whole range of possibilities and react accordingly'' \cite{Watts2012}. In the present context, this strategy implies we can create better algorithms for option ranking by observing, and addressing, how people respond to social influence and position biases. We show this strategy applied in RAICR improves upon simple algorithms used in the past, which include ordering results by popularity \cite{Burghardt2018} or recency \cite{lerman14as}. While some crowdsourced ranking strategies use a two-tier platform model, in which researchers rank options based on whether content is downloaded and rated \cite{Salganik2006,Antenore2018,Abeliuk2017www}, RAICR is based on a common simpler model in which we only observe if content is chosen \cite{Burghardt2018,PopularDynam}. This subtle difference implies many previous ranking schemes are not applicable. The present paper also compliments previous work that uses features, such as answer position, to predict Q\&A website quality \cite{Shah2010,MyopiaCrowd}.

\section{Experiment}

The experiment asked subjects, hired through Amazon Mechanical Turk between August 2018 and September 2019, to answer a series of questions shown in the Appendix. The order of questions was randomized for each subject, and questions were not time-limited. Questions include specifying the ratio of the areas of two shapes or the lengths of two lines, or the number of dots in an image. We designed the experiment around quotidian tasks that do not require specific expertise but are difficult for most people. Despite their difficulty, the questions have objectively correct answers. We quantify the quality of an answer by the difference from the mean of all guesses, which better matches a log-normal distribution, as discussed later. In the Appendix, we define quality of an answer by the difference from the correct value and find results are qualitatively very similar. 

Subjects were assigned to one of three conditions. In the \textit{guess condition}, shown in the left panel of Fig.~\ref{fig:ExperimentSchematics}, the subjects could freely type their guesses. In the \textit{control condition}, shown in the center panel of Fig.~\ref{fig:ExperimentSchematics}, subjects were told to choose the best among two options, and were not told how the two answers were ordered. Finally, in the \textit{social influence condition}, shown in the right panel of Fig.~\ref{fig:ExperimentSchematics}, they were told the first among two answers was the most popular, but the layout was otherwise identical to the previous condition. We only showed ten questions to each subject to reduce the effects of performance depletion in Q\&A systems~\cite{Ferrara17}. Further, to reduce bias due to other phenomena, such as the decoy effect \cite{Huber1982}, we showed subjects only two choices. Subjects in the same condition but assigned on different days have statistically similar behavior.

Approximately 1800 subjects are evenly split between the conditions (596, 586, and 587 for the guess, control, and social influence condition, respectively). For guess values, we remove extreme outliers (guesses smaller than 1 or greater than $10^6$). All answers 
are supposed to be greater than 1, while values greater than $10^6$ may affect mean values and appear to represent throwaway answers. The number of valid participants for each question is shown in Table\ref{tab:guesssummary}.

\begin{table}[t]
\centering
\caption{Number of subjects for each guess question (after cleaning)}
\label{tab:guesssummary}
\begin{tabular}{|c|c|c|c|c|c|c|c|c|c|}
\hline
Q1&Q2&Q3&Q4&Q5&Q6&Q7&Q8&Q9&Q10\\ \hline
595&594&593&595&592&593&589&591&584&590\\ \hline
\end{tabular}
\begin{flushleft}
\end{flushleft}
\end{table}

Mechanical Turk workers were hired if they had an approval rate of over $95\%$, completed more than 1000 Human Intelligence Tasks (HITs), and never participated in any of the experiment conditions before. Each worker was paid $\$1.00$ for the guessing condition and $\$0.50$ for the other two conditions. The assignment took $6$ minutes on average for the guess condition and $2$ minutes on average for the other conditions, equivalent to an hourly wage of $\$10-\$15$. The human experiment was approved by the appropriate IRB board.

\section{Results}
\subsection{A Mathematical Model of Decisions}

\begin{figure}[tbh!]
\centering
\includegraphics[width=0.7\textwidth]{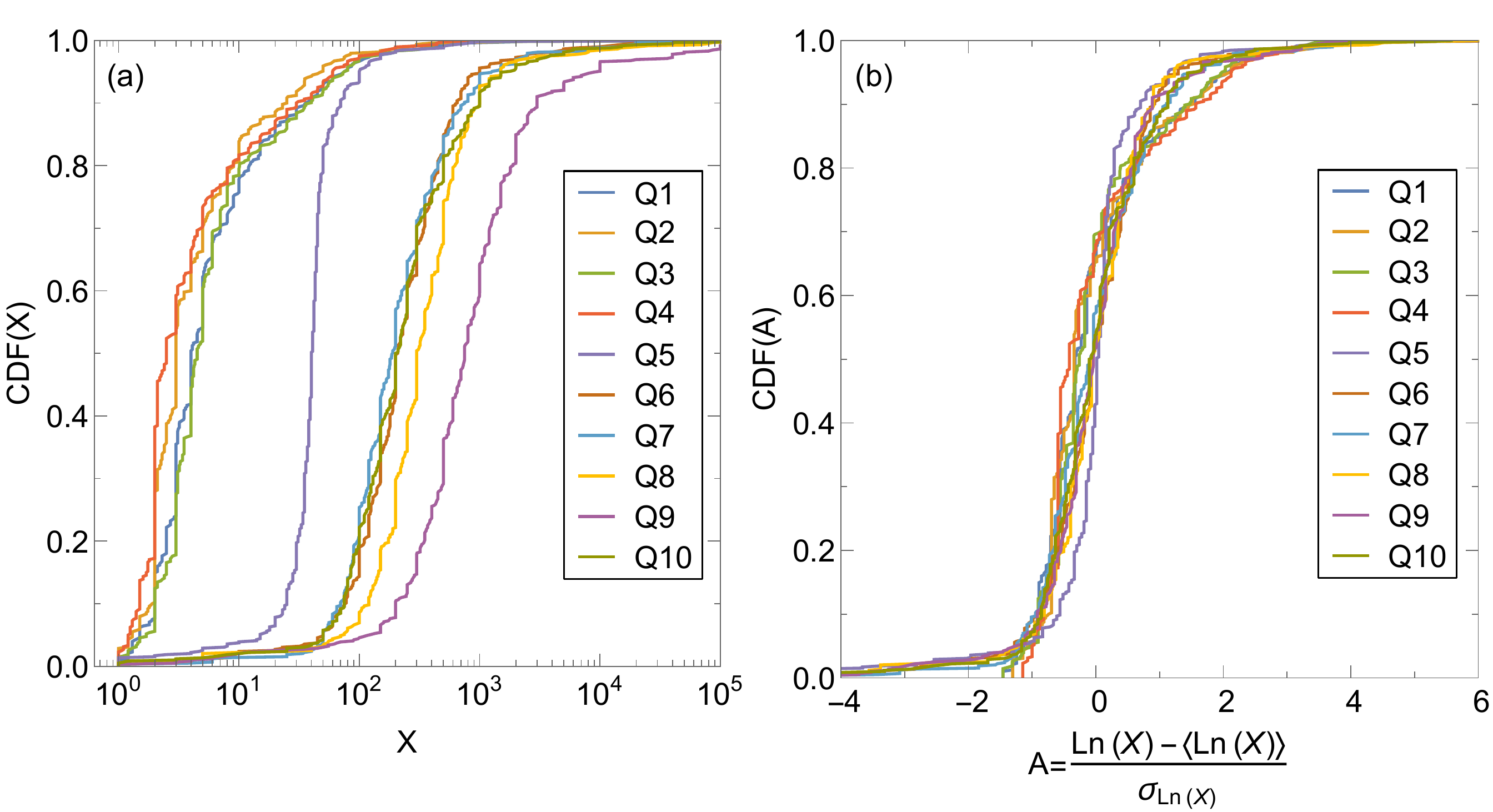}
\caption{\label{fig:CDFGuess} The CDF of guesses for each question. (a) Before normalization, and (b) after normalization.
}
\end{figure}

We use data gathered from the experiment to answer \textbf{RQ1}: \textit{How does quality and presentation of options jointly impact individual decisions?}

\subsubsection{Open-Ended Experiment Condition} 
To derive the model of how people make these decisions, we start by constructing the distribution of guesses for each question (the \textit{guess} condition in the experiment). 
The guesses, plotted in Appendix Fig.~\ref{fig:GuessExperimentResults} and CDF shown in Fig.~\ref{fig:CDFGuess}a, are highly variable (by as many as six orders of magnitude), while the correct answers vary by three orders of magnitude. The median guess may differ from the true answer, in agreement with previous work \cite{Kao2018}, but values are typically the correct order of magnitude.

We normalize guesses by defining a new variable $A$:
\begin{equation*}
    A = \frac{\text{ln}(X)-\langle \text{ln}(X)\rangle}{\sigma_{\text{ln}(X)}},
\end{equation*}
where $X$ is a guess value, and $\langle \text{ln}(X)\rangle$ is the mean of the logarithm of guesses. Figure~\ref{fig:CDFGuess}b shows that this simple normalization scheme effectively collapses answer guesses to a single distribution. We show in the Appendix that guesses are not normally distributed, but instead are better approximated as log-normal, in agreement with previous work on a different set of questions \cite{Kao2018}. The normalized guesses $A$ can be thought of as the $z$-scores in log-normal distributions. Alternative ways to center data, shown in the Appendix, produce similar results. We use these normalized guesses and distributions in the remaining two experiment conditions. Intuitively, if the mean of all guesses converges to the correct answer, $A=0$ can be thought of as the best answer.

\subsubsection{Two-Choice Experiment Conditions}
\begin{figure}[tbh!]
\centering
\includegraphics[width=0.3\textwidth]{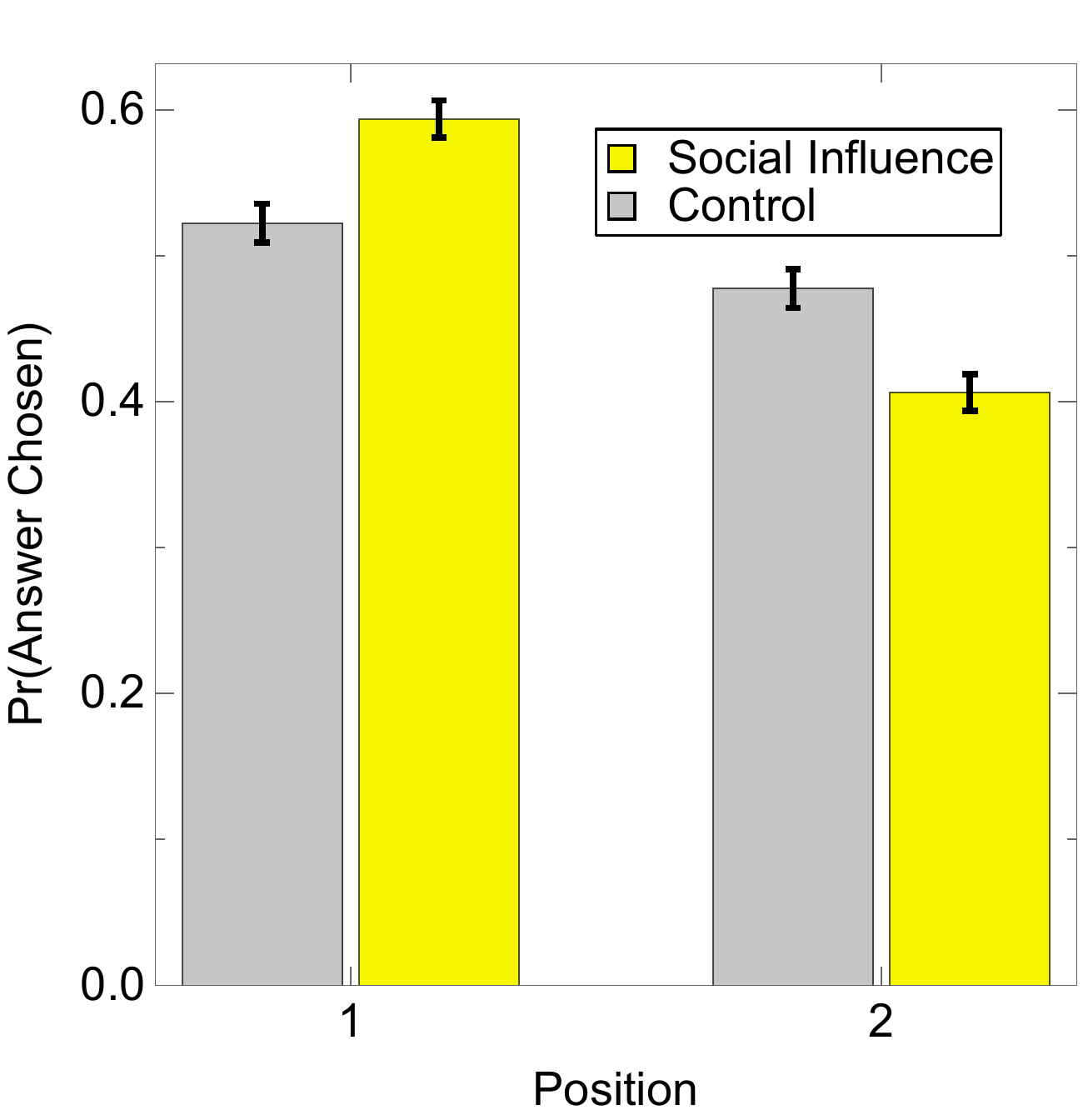}
\caption{\label{fig:ChooseVsPos} Probability to choose an answer versus its position for the social influence and control condition.
}
\end{figure}

The latter two experiment conditions require subjects to pick the best among two answers to the question. For both the control and social influence conditions, answers are ordered vertically, with one answer above the other. There is a significant position effect when answers are ordered this way, as shown in Fig.~\ref{fig:ChooseVsPos}. In the control condition, there is a slightly greater probability (52\%) to choose the first (top) answer over the last one (p-value $<0.001$). In the social influence condition, meanwhile, the probability to choose the first answer is substantially larger (59\%) and statistically significantly different than the control condition (p-value $<0.001$). This is in agreement with previous work showing that social influence amplifies the position effect~\cite{Burghardt2018}.

\subsubsection{The Biased Initial Guess Model}
\begin{figure}[tbh!]
\centering
\includegraphics[width=0.4\textwidth]{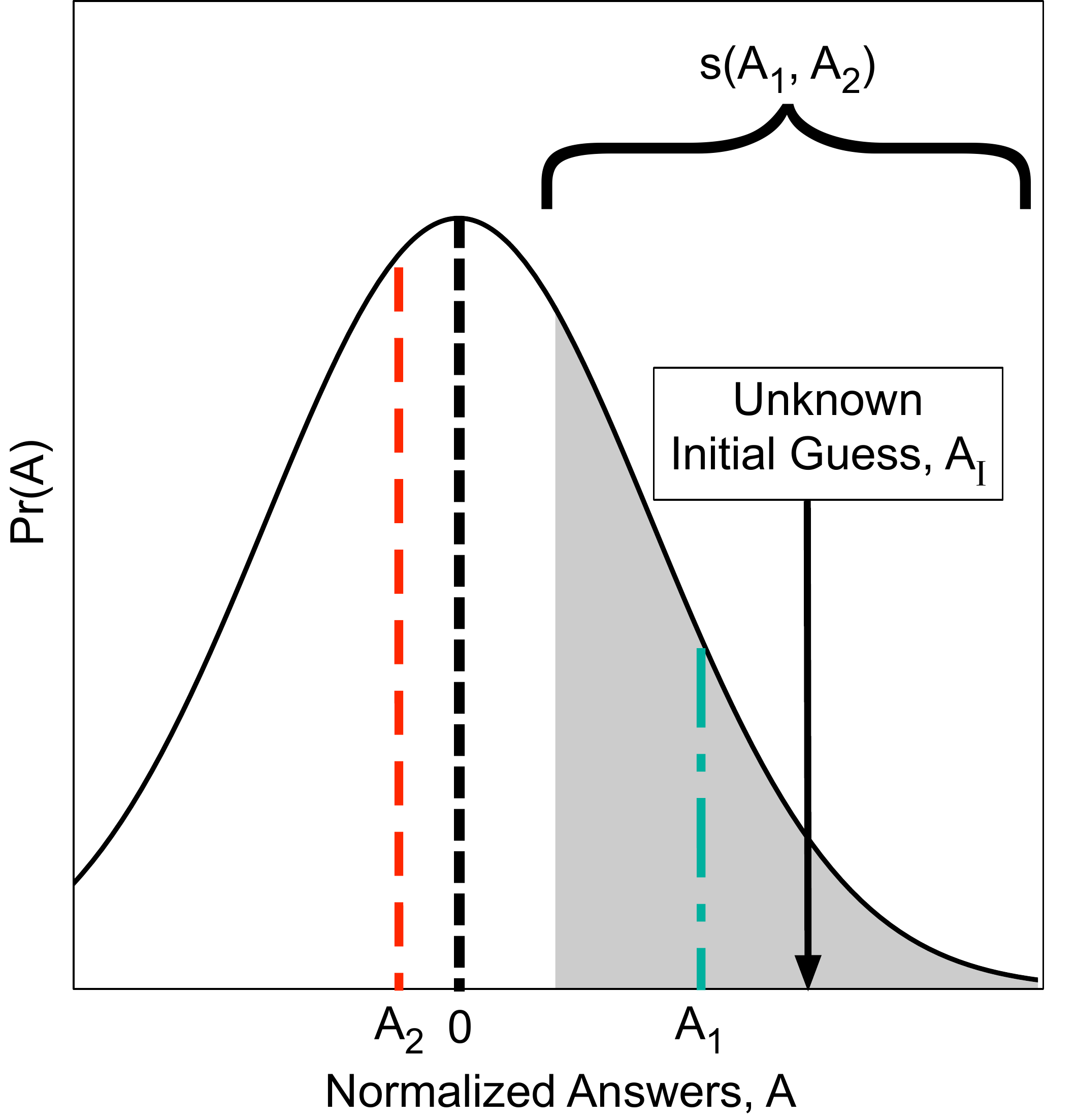}
\caption{\label{fig:ModelSchematics} Schematic to calculate $s(A_1,A_2)$. Guesses follow an approximately log-normal distribution unique to each question, but the normalized guesses, $A$ are approximately standard normal distributed. In an experiment, assume two candidate answers are provided, $A_1$ is listed first, and $A_2$ is listed last. The variable $s(A_1,A_2)$ in Eq.~\ref{eq:initselect} is the probability an unknown initial guess is closer to $A_1$ than $A_2$.
}
\end{figure}

We now have the necessary ingredients to model how decisions to choose an answer are affected by its quality, position, and social influence. 
We present the Biased Initial Guess (BIG) decision model and show it is consistent with the data.

We first discuss the simplest case where a user has to choose the better of two answers $A_1$, listed first, and $A_2$, listed last, in the absence of cognitive biases. Figure~\ref{fig:ModelSchematics} shows probability of the answer, with the normalized values of the choices $A_1$ and $A_2$, as well as the user's initial guess $A_I$ about the true answer, which we do not observe. All things equal, the user will choose the first answer $A_1$ if it is closer to the initial guess, i.e., if $|A_1-A_I|<|A_2-A_I|$, and will otherwise choose $A_2$. The probability to choose $A_1$ is then:
\begin{equation}
s(A_1,A_2) = 
\begin{cases}
\text{Pr}(A_I > \frac{A_1+A_2}{2}) & A_1 > A_2 \\
1/2 & A_1 = A_2\\
\text{Pr}(A_I < \frac{A_1+A_2}{2}) & A_1 < A_2
\end{cases}
\end{equation}
 Assuming $A_1$ and $A_2$ follow a normal distribution quantified in the guess experiment condition,
\begin{equation}
\label{eq:initselect}
s(A_1,A_2)= 
\begin{cases}
\text{erfc}\left(\frac{1}{\sqrt 2}\frac{A_1+A_2}{2}\right)/2 & A_1 \ge A_2 \\
\left[1 + \text{erf}\left(\frac{1}{\sqrt 2}\frac{A_1+A_2}{2}\right)\right]/2, & A_1 < A_2
\end{cases}
\end{equation}
where $\text{erf}(.)$ is the error function, $\text{erfc}(.) = 1-\text{erf}(.)$ is the complimentary error function. When $s(A_1,A_2)>0.5$, $A_1$ is not only more likely to be closer to the initial guess ($A_I$) than $A_2$, but is also closer to zero than $A_2$, and therefore the objectively better answer. On the other hand, when $s(A_1,A_2)<0.5$, $A_1$ is further from most guesses compared to $A_2$ and the objectively worse answer.

To better model decision-making, we have to account for biases, due to cognitive heuristics and algorithmic ranking, to explain why people do not always choose the best option. As shown in Fig.~\ref{fig:ChooseVsPos}, sometimes they choose the first answer even if it is not the best answer. We quantify this position bias by assuming that with probability $p$ participants choose the first answer regardless of its quality. This parameter should presumably be small in the control condition and large in the social influence condition. Subjects may also choose an answer regardless of its position or quality because there is no monetary incentive to choose good answers. We model this by allowing subjects to choose an answer at random with a probability $r$. Taking these two heuristics into account, we arrive at the BIG Model: 

\begin{equation}
\label{eq:model_full}
\begin{split}
\text{Pr}(\text{Choose}~A_{1}) &= r/2 + (1 - r) [p + (1 - p) s(A_1,A_2)]\\
&= \begin{cases}
r/2 + (1 - r) \left[p + (1 - p)\text{erfc}\left(\frac{1}{\sqrt 2}\frac{A_1+A_2}{2}\right)/2\right] & A_1 \ge A_2 \\[7pt]
r/2 + (1 - r) \left[p + (1 - p)\left[1 + \text{erf}\left(\frac{1}{\sqrt 2}\frac{A_1+A_2}{2}\right)\right]/2\right], & A_1 < A_2
\end{cases}
\end{split}
\end{equation}
The probability of choosing $A_2$ is simply the compliment of this probability. Because $A=0$ is expected to be the best answer, with some simple manipulation we can infer the probability the best answer is chosen.

\begin{equation}
\label{eq:model_best}
\text{Pr}(\text{Choose}~A_{1}|A_{1}~\text{Best}) = \begin{cases}
r/2 + (1 - r) \left[p + (1 - p) \text{erfc}\left(\frac{1}{\sqrt 2}\frac{A_1+A_2}{2}\right)/2\right]  & \frac{A_1+A_2}{2}<0\\[7pt]
r/2 + (1 - r) \left[p + (1 - p) \left(1-\text{erfc}\left(\frac{1}{\sqrt 2}\frac{A_1+A_2}{2}\right)/2\right)\right] & \frac{A_1+A_2}{2}\ge0\\
\end{cases}
\end{equation}
A similar equation can model $\text{Pr}(\text{Choose}~A_{2}|A_{2}~\text{Best})$.

\begin{figure}[tbh!]
\centering
\includegraphics[width=0.8\textwidth]{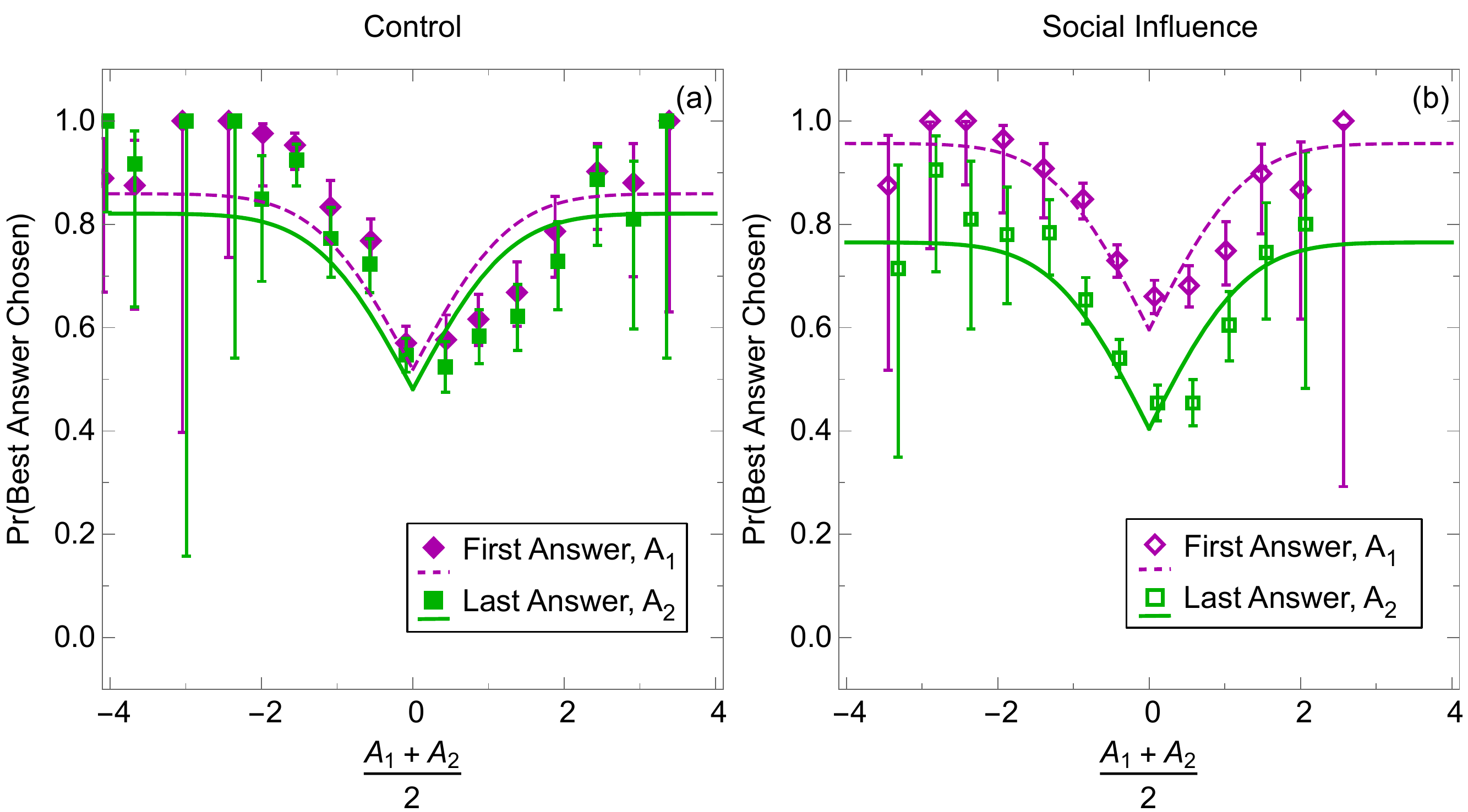}
\caption{\label{fig:ModelFit} Decision model agreement with experiment data. Plots show the probability the best answer is chosen for the model (lines) and experiment (symbols) under (a) the control condition and (b) the social influence condition. Purple diamonds are probabilities when the best answer is the first answer, $A_1$ and green squares for when the best answer is $A_2$.
}
\end{figure}

Agreement between the model and data is shown in Fig.~\ref{fig:ModelFit}. In the \textit{control condition} (Fig.~\ref{fig:ModelFit}a), the best parameters are $r=0.28\pm0.02$ and $p=0.05\pm0.02$. We find that the log-likelihood of the model, $\ell=-3002.49$, is not statistically different from log-likelihood if the data came from the model itself: p-value $=0.40$. See Methods for how p-values and error bars are calculated. We also check if we need both parameters, $r$ and $p$, using the likelihood ratio test and Wilks' Theorem~\cite{Wilks1938}. We compare the likelihood ratio of the two-parameter model to simpler models with $p$ or $r$ (or both) set to zero. The probability a simpler model could fit the data as well or better is $\le 0.002$. We conclude that our model describes the control condition very well. 

The agreement between data and model is similarly close in the \textit{social influence condition} (Fig.~\ref{fig:ModelFit}b). The position bias parameter $p=0.21\pm 0.01$ is larger than in the control condition, in agreement with expectations. We also find $r=0.08\pm 0.02$, thus social influence reduces the frequency of random guesses. In both experiment conditions, surprisingly, $\approx 20\%$ of users choose answers for reasons besides ``quality'' ($r/2+(1-r)p = 18\%$ and $23\%$ for the control and social influence conditions, respectively). Similar to the control condition, we find that the model is consistent with the data. The log-likelihood of the empirical data ($\ell=-3190.13$) is not statistically different from log-likelihood values if the data came from the model: p-value $0.47$. The probability a simpler model ($r$ or $p$ set to zero) could fit the data as well or better is $<10^{-5}$. In conclusion, we find the BIG model is consistent with both experiment conditions and its parameters are interpretable and meaningful. In the Appendix, we show that all these results are consistent when we look at a subset of experiment questions or center the data differently. 

\subsection{Algorithmic Ranking Instability}
Crowdsourcing websites automatically highlight what they consider the best choices to help their users more quickly discover them. For example, Stack Exchange (like other Q\&A platforms) usually ranks answers to questions by the number of votes they receive. Despite problems with popularity-based ranking identified in previous studies~\cite{Salganik2006,lerman14as}, it is widely used for ranking content in crowdsourcing websites. In this section we identify an instability in popularity-based ranking: the first few votes a worse answer receives can lock it in the top position, where it acquires cumulative advantage \cite{Rijt2014}. This allows us to answer \textit{RQ2: When is algorithmic ranking unstable and does not reliably identify the best option?}

\begin{figure*}[tbh!]
\centering
\includegraphics[width=0.95\textwidth]{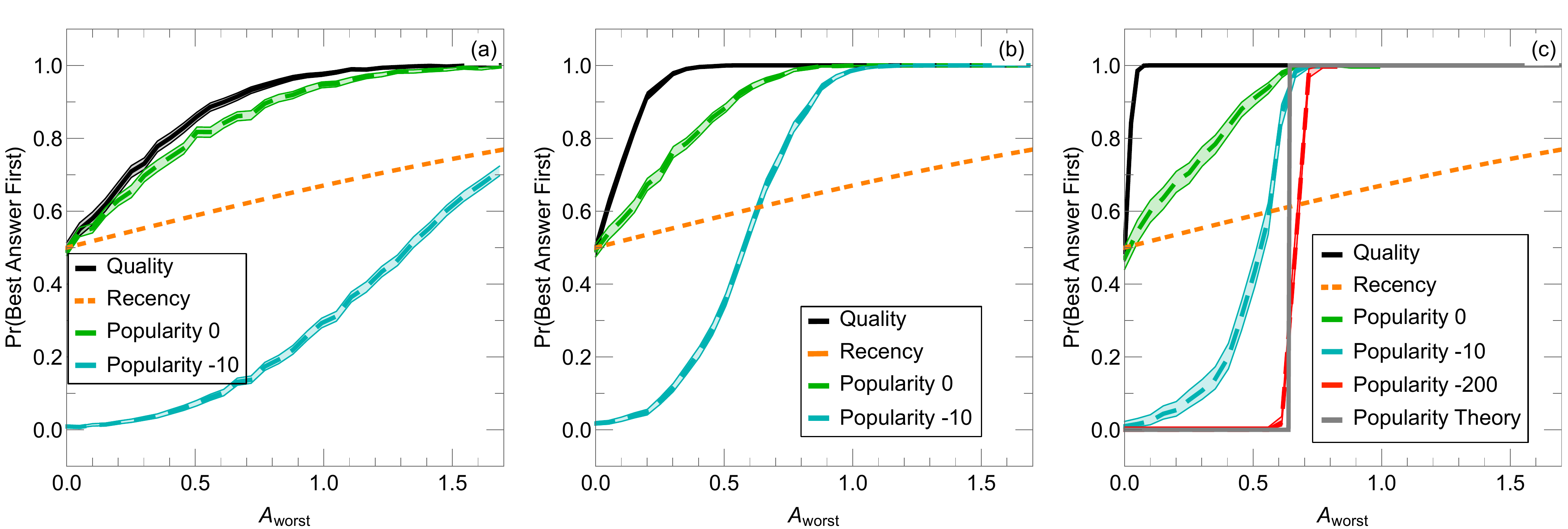}
\caption{Comparison of ranking policies via simulations. Plots show the probability the best answer is ranked first after (a) 50, (b) 500, and (c) 20,000 votes, when answers are ranked by quality (black line), recency (orange dashed line), and popularity. Also shown in (c) is the critical value of $A_{\text{worst}}$ based on Eqs.~\ref{eq:scrit} and~\ref{eq:initselect}. In these simulations, subjects choose answers following the BIG Model with $p=0.2$ and $r=0.09$. ``Popularity 0'' (green dashed line), ``Popularity -10'' (cyan dashed line), and ``Popularity -200'' (red dashed line) means that the worst answer starts with a 0, 10, or 200 vote advantage, respectively. Shaded areas are 95\% confidence intervals. 
\label{fig:InitialGuessModelRankData} 
}
\end{figure*}

To demonstrate the instability, we simulate a group of agents who choose answers according to the BIG model with $p=0.2$ and $r=0.09$. In the simulations, one answer is objectively best, e.g., exactly equal to the correct answer ($A_{\text{best}}=0$), while the worst answer is larger than $A_{\text{best}}$ (results are symmetric if $A_{\text{worst}}<A_{\text{best}}$). At each timestep, a new user arrives and independently chooses the first or last answer according to Eq.~\ref{eq:model_full}. Answers reorder depending on the ranking algorithm. Figure~\ref{fig:InitialGuessModelRankData} shows our results. If the worst answer, $A_{\text{worst}}$, is not too large and has a few extra votes (a common occurrence when the worst answer is posted before the better answer), we find that popularity ordering completely breaks down---the worst answer usually becomes more popular and is ranked first. Even when both answers start with the same number of votes, the worse answer is often ranked first. Results remain stable even after 20K votes; the effect does not appear transient.

\begin{figure}[tbh!]
\centering
\includegraphics[width=0.44\textwidth]{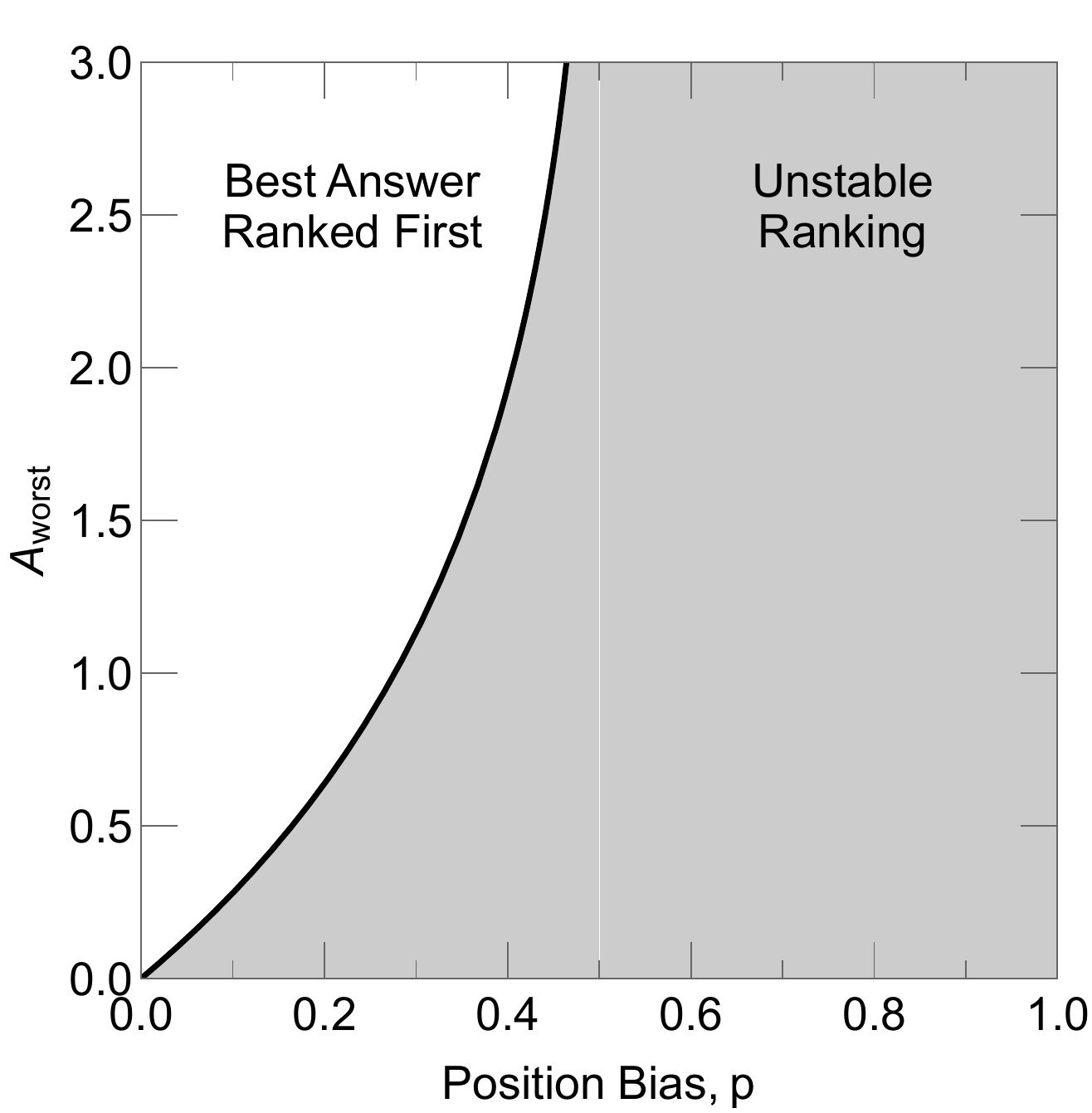}
\caption{\label{fig:SimPhaseSpace} Phase diagram for popularity-based ranking based on Eqs.~\ref{eq:scrit} and~\ref{eq:initselect}. We show the boundary between regimes where best answers are guaranteed to be ranked first with enough votes (white area) and where ranking is unstable (gray area) as a function of position bias ($p$) and the value of the worse answer ($A_{\text{worst}}$). In this plot, larger values of $A_{\text{worst}}>0$ correspond to worse answers and results are symmetric for $A_{\text{worst}}<0$. 
}
\end{figure}

We can explain this result using our model. Using the compliment of Eq.~\ref{eq:model_full}, we can define the probability the best answer is chosen, conditional on it being ranked last:
\begin{equation}
    \text{Pr}(\text{Choose}~ A_{\text{best}}=A_2) = r/2 + (1 - r) (1 - p) s(A_1,A_2)
\end{equation}
We find that
\begin{equation}
    \text{Pr}(\text{Choose}~ A_{\text{best}}=A_2) = \text{Pr}(\text{Choose}~ A_{\text{best}}=A_1)-p(1-r)
\end{equation}
where $\text{Pr}(\text{Choose}~ A_{\text{best}}=A_1)$ is the probability of choosing the best answer when it is ranked first. If \begin{equation}
    \text{Pr}(\text{Choose}~ A_{\text{best}}=A_2)>\text{Pr}(\text{Choose}~ A_{\text{worst}}=A_1)
\end{equation} 
then subjects are more likely to choose the better answer regardless of whether its ranked first or second, therefore answer order is stable. When the above inequality is not true, however, then subjects are more likely to pick the first answer regardless of its quality, and the popularity ranking is unstable. The critical value of $A_{\text{worst}}$ between the two regimes is when 
\begin{equation}\label{eq:scrit}
    s_{\text{crit}}=\frac{1}{2 (1-p)}
\end{equation}
and is independent of $r$. Intuitively, if the answer is exceptionally bad, it will always be less popular. This is alike to the results in Salganik's MusicLab study \cite{Salganik2006}, where particularly good and bad songs were ranked correctly. However, if the first answer is likely to be chosen regardless of quality, the worst answer will continue accumulating votes and remain in the top position. Given $A_{\text{best}}=0$, we can use Eqs.~\ref{eq:scrit} and~\ref{eq:initselect} to numerically solve for the critical value of $A_{\text{worst}}$. We plot the critical point as a function of $p$ and $A_{\text{worst}}$ in phase diagram in Fig.~\ref{fig:SimPhaseSpace}. We see that there is always a large part of the phase space where popularity-based ranking will be unpredictable if answer quality is close together. Based on Eq.~\ref{eq:scrit}, if $p\ge 0.5$, there will be no case where popularity-based ranking is guaranteed to correctly rank answers. While this is an extreme case, it still points to substantial limitations of popularity-based ranking.

\subsection{Stabilizing Algorithmic Ranking}

Given the problems with popularity ranking, it is critical to create a more consistent ranking algorithm. Moreover, while we have so-far explored questions with numeric answers, we want a method that works for all types of answers. Using the BIG model, we can answer RQ3: \textit{How can we stabilize algorithmic ranking such that the best option is typically ranked first?}

Assume we can approximate $p$ and $r$, then we can invert Eq.~\ref{eq:model_full} and use votes to infer the only unknown variable $s(A_{\text{best}},A_{\text{worst}})$. When $s(A_{\text{best}},A_{\text{worst}})>0.5$, $A_{\text{best}}$ is the best answer, but if we incorrectly rank $A_{\text{worst}}$ first, $s(A_{\text{worst}},A_{\text{best}})<0.5$. We can therefore rank the answer in which $s(A_{\text{best}},A_{\text{worst}})>0.5$ as the best answer. This is the backbone of the RAICR algorithm. The algorithm uses maximum likelihood estimation to solve for $s(A_{\text{best}},A_{\text{worst}})$, as shown in the Appendix. Therefore, if the data matches the BIG model with the correct $r$ and $p$ parameters, our method \emph{optimally infers the correct raking} by having minimal variance and no bias \cite{Newey1994}. Moreover, \emph{this method only depends on the votes an answer receives rather than the type of answer}, such as a numerical or textual answer. We compare quality ranking to popularity-based ranking, and recency-based ranking (ranking by the last answer picked), as shown in Fig.~\ref{fig:InitialGuessModelRankData}. The probability recency ranks the best answer first is calculated as the self-consistent equation: 
\begin{dmath}
    \text{Pr}(\text{Rank $A_{\text{best}}$ First}|\text{Recency}) = \text{Pr}(\text{Choose}~ A_{\text{best}}|A_{\text{best}}~\text{First}) \text{Pr}(\text{Rank $A_{\text{best}}$ First}|\text{Recency})\\  +  \text{Pr}(\text{Choose}~ A_{\text{best}}|A_{\text{best}}~\text{Last})(1 - \text{Pr}(\text{Rank $A_{\text{best}}$ First}|\text{Recency}) ).
\end{dmath}
This represents the limit answers acquire many votes. The solution to this equation is:
\begin{dmath}
    \text{Pr}(\text{Rank $A_{\text{best}}$ First}|\text{Recency}) = \frac{2 (1-p) (1-r) s(A_{\text{best}},A_{\text{worst}})
    +r}{2-2 p (1-r)}
\end{dmath}
We find that the RAICR algorithm performs at least as well as popularity-based ranking by ranking the better answer first, and often better after 20-50 votes (Fig.~\ref{fig:InitialGuessModelRankData}a and Appendix Fig.~\ref{fig:SimPRobust}). Moreover, RAICR always outperforms the recency-based algorithm. The benefit of the RAICR algorithm only improves as we collect more votes. 
For example, Fig.~\ref{fig:InitialGuessModelRankData}b shows after after 500 votes the advantage of RAICR is larger, and Fig.~\ref{fig:InitialGuessModelRankData}c shows that after 20K votes the method is nearly optimal. Popularity-based ranking performs badly for $A_{worse}<0.6$, and recency performs worst if $A_{worse}>0.6$. 

While we show these results hold when we have 20 to 20,0000 votes, many platforms are underprovisioned, with  a large fraction of webpages 
receiving little attention and votes~\cite{baeza2018bias,Gilbert2013}. It is the webpages that receive many votes, however, which may be the most important. Correctly ranking options in these popular pages is therefore especially critical to crowdsourcing websites. Moreover, a moderate number of votes is generally needed to make a reasonable estimate of quality, so very few methods will accurately rank unpopular pages.

One caveat of the RAICR algorithm is that we need an approximate value for two parameters: $r$ and $p$. What happens if either of these parameters are far off? For example, we could assume $r=0$ when $r=0.09$. Results in the Appendix, however, show that the findings are quantitatively very similar, and therefore our model is robust to assumptions about $r$. On the other hand, what if we incorrectly estimate both $r$ and $p$? We show in the Appendix that even in this worst-case scenario, our method performs slightly worse, but is comparable or substantially better than popularity-based ranking. 

\section{Future Work and Design Implications}

In the experiment, we decomposed the crowdsourcing task to its basic components to reduce the complexity and variation inherent in real world tasks. While this helps to disentangle effects of option quality and position without confounding factors muddying the relationships, future work is needed to verify ecological validity of our results \cite{ObservationalVsExp}. It is encouraging that the probability to choose an answer (Fig.~\ref{fig:ChooseVsPos}b) is quantitatively similar to empirical data gathered from Stack Exchange \cite{Burghardt2018}, despite Mechanical Turk workers not being representative of the general population \cite{Munger2019}. 

For simplicity, we only explored two-option questions; future work should aim to understand multi-option decision-making given options of variable quality. A generalization of RAICR should also address more complicated biases, such as preference for round numbers \cite{Fitz1986}, anchoring \cite{shokouhi2015anchoring,Furnham2011}, and biases that appear in multi-option decisions, such as the decoy effect \cite{Huber1982}. Finally, while RAICR is found to be robust to moderate changes in its parameters, this algorithm and its extensions may fail to rank options properly if its parameters are far off, or if the BIG model is wrong. In the experiment, the model is backed by data, but future work needs to address whether other tasks or questions follow this model. 

Our results offer implications for crowdsourcing platforms. First, designers must recognize the limits of crowdsourcing due to biases implicit in their platform. In our experiment people often upvoted options at random (up to 20\% of all votes), and chose an inferior option simply because it was shown first. This creates a ranking instability when options are of similar quality. Our controlled experiment and mathematical model point to ways we can counteract this instability. Designers should similarly create platform-tailored mathematical models and controlled experiments to rigorously test how crowds can better infer the best options. 

A key property of our RAICR algorithm is that it relies on accurate modeling of user decisions to counteract cognitive biases. In effect, each vote is weighted depending on the ranks of answers at the time the vote is cast. The idea is similar to one described by Abeliuk et al.~\citeyear{Abeliuk2017www} that ranks items by their inferred quality in order to more robustly identify blockbuster items. Similar weighting schemes could be applied to future debiased algorithms to address the unique goals of each crowdsourcing platform. 

There are also simple methods that platforms like  Reddit, Facebook, and Stack Exchange can try that may greatly outperform the baselines we mention in our paper. For example, items that have not yet acquired many votes can be ranked randomly to reduce initial ranking biases. Alternatively, new posts and links could be ranked appropriately but their popularity could be hidden until they gather enough votes. Our results suggest this could reduce social influence-based position bias up until the true option quality is more obvious. 

\section{Conclusion}
In this paper, we introduce an experiment designed to inform how cognitive biases and option quality interact to affect crowdsourced ranking. Results from this experiment help us create a novel mathematical decision model, the BIG model, that greatly improves our understanding of how people find the best answer to a question as a function of answer quality, rank, and social influence. This model is then applied to the RAICR algorithm to better rank answers. The BIG model also helped us uncover instability in popularity-based ranking. The instability depends on the quality of options: when there are large differences between option qualities, popularity converges optimally and predictably. However, when the difference between the quality of options is small, the better option may not always become the most popular. These results can help us better understand the foundational empirical results of Salganik et al. (\citeyear{Salganik2006}), who found that popularity-based ranking correctly ranked high and low quality songs, while the ranking of intermediate quality songs was highly unstable. Although our experimental setup is undeniably simpler than real crowdsourcing websites, our results suggest that accurate models of user behavior together with mathematically principled inference can improve the efficiency of crowdsourcing. 


\begin{acks}
Our work is supported by the US Army Research Office MURI Award No. W911NF-13-1-0340 and the DARPA Award No. W911NF-17-1-0077. Data as well as code to create experiments, create simulations, and analyze data is available at https://github.com/KeithBurghardt/QualityRankCodeAndData.
\end{acks}

\appendix
\section{Appendix}
In this section, we discuss the experiment questions, the validity of the log-normal guess distribution, alternative answer normalization schemes, and the log-likelihood estimate used to rank answers in simulations. We also discuss the robustness of the simulation results.

\subsection{Experiment Details}

\begin{figure*}[tbh!]
\centering
\includegraphics[width=1\textwidth]{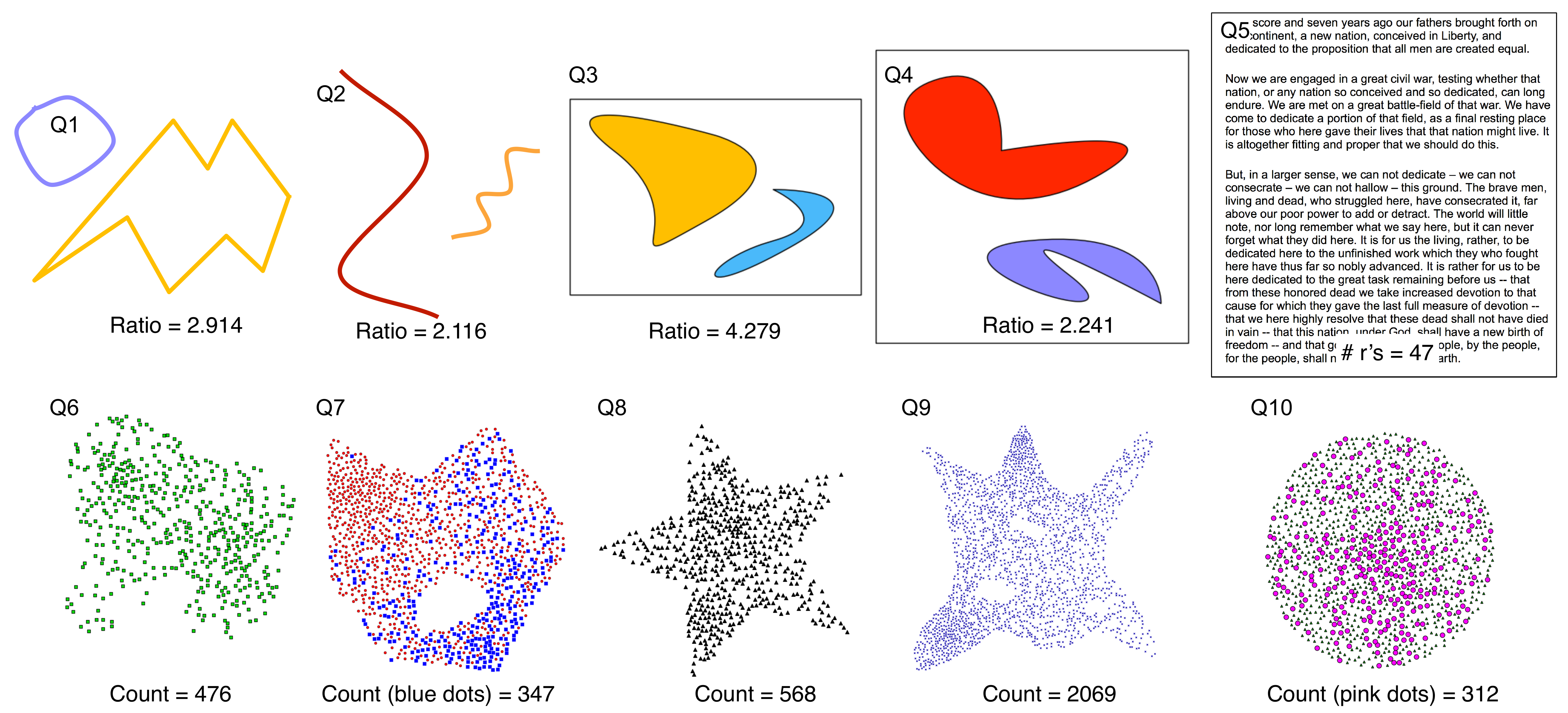}
\caption{\label{fig:Questions} Figures associated with the questions used in our study. Questions were: (Q1) find the perimeter ratio of the larger to smaller shape, (Q2) find the length ratio of the larger to smaller line, (Q3--Q4) find the area ratio of the larger to smaller shape, (Q5) find the number of ``r'''s in Lincoln's Ghettysburg Address, (Q6--Q10) find the number of dots. For Q7 and Q10 we specify the color of the dots to count. Correct answers to these questions are listed below the figure.
}
\end{figure*}

\begin{figure}[tbh!]
\centering
\includegraphics[width=0.4\textwidth]{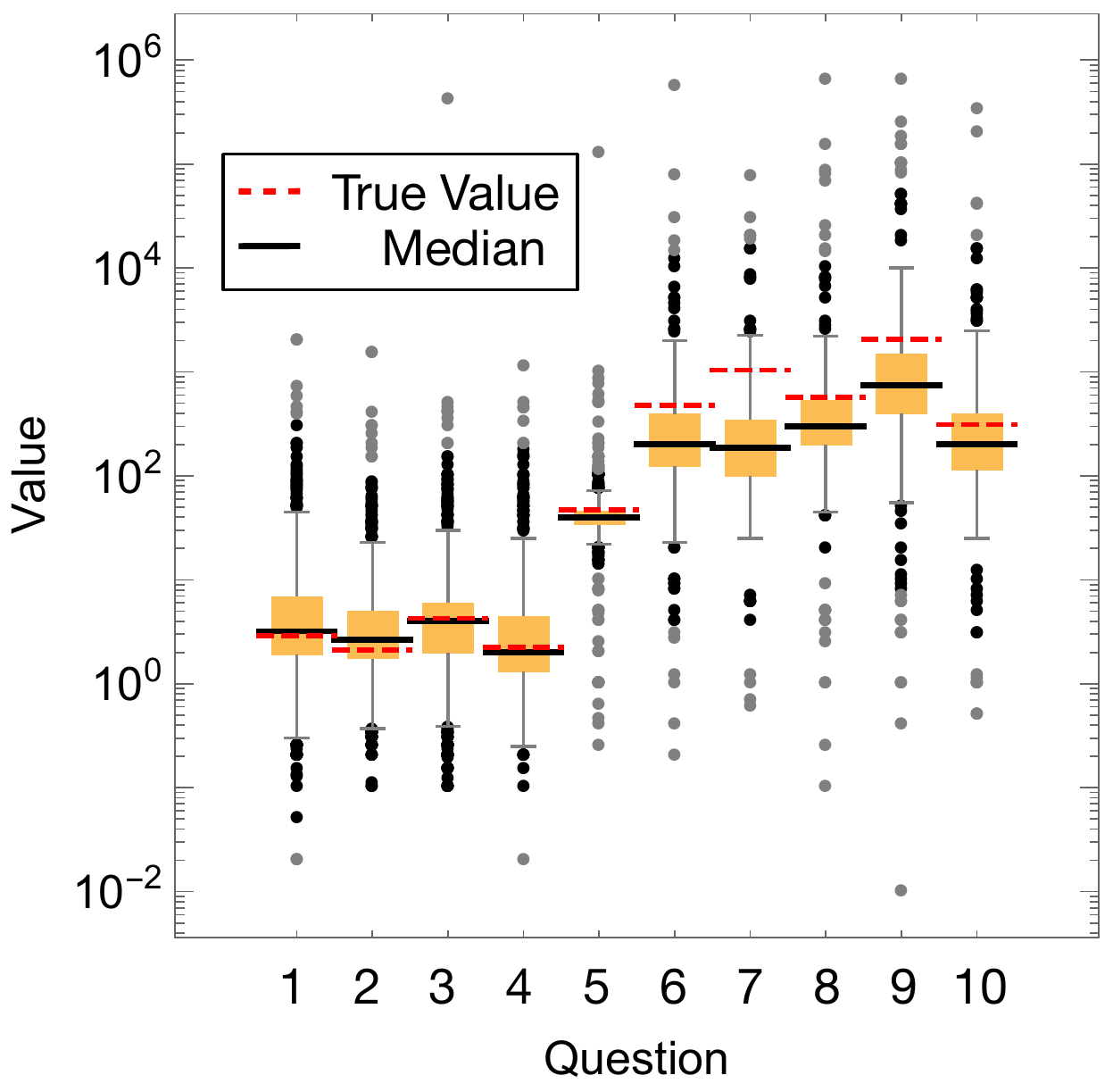}
\caption{\label{fig:GuessExperimentResults} Guesses for each question. Black line: median guess, red dashed line: true value. We see the median guess are typically close to the true value.
}
\end{figure}

\begin{figure*}[tbh!]
\centering
\includegraphics[width=1\textwidth]{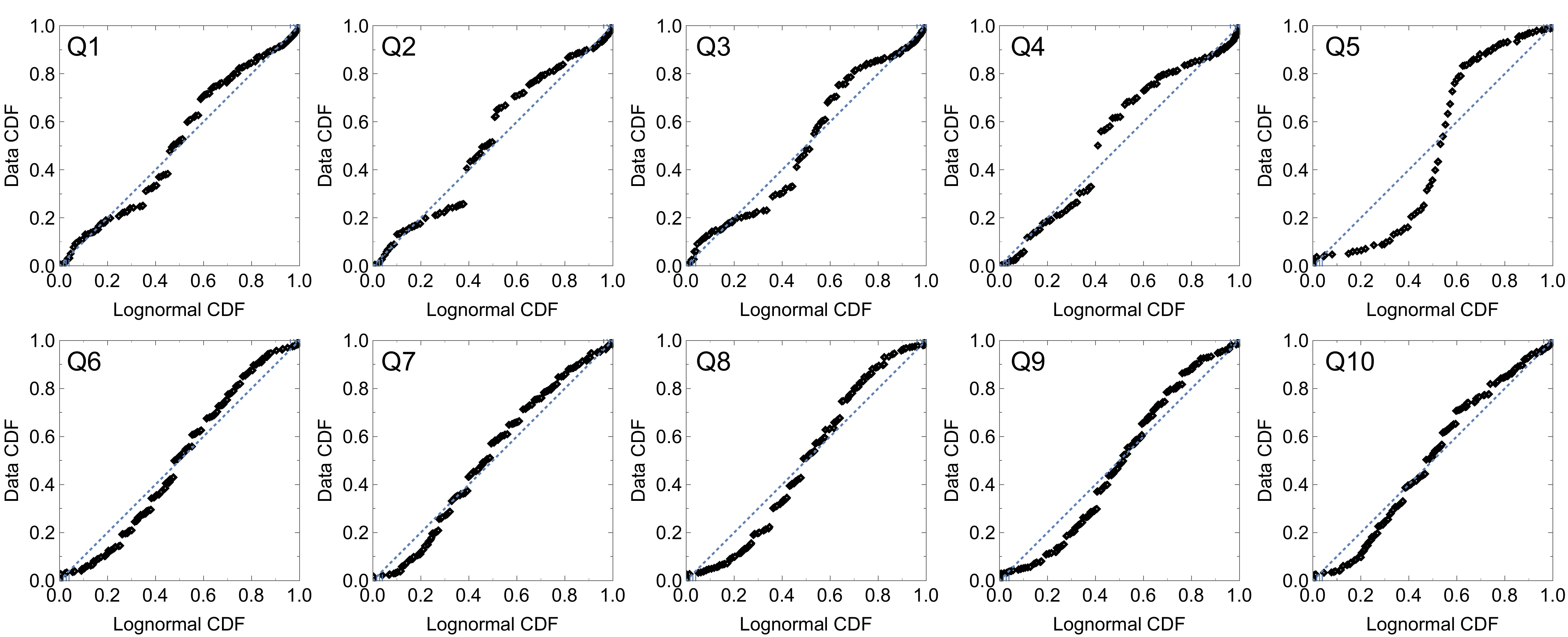}
\caption{\label{fig:QQplots} Quantile-quantile plots comparing empirical data to a log-normal distribution. We see agreement is reasonable except for Q5, and is very good for Q6--10. Q5 corresponds to guessing the number of ``r''s in Lincoln's Ghettysburg Address, which, unlike other questions, can be counted directly.
}
\end{figure*}

Experiment questions are shown in Fig.~\ref{fig:Questions}. We see that questions cover a variety of visual problems with numerical answers. Questions include finding ratios of lines or areas, counting the number of ``r''s in text, or counting dots. We make sure that guesses cannot be easily measured (e.g., lines are not straight and dots are not evenly distributed). The guesses, median guess value, and true values are shown in Fig.~\ref{fig:GuessExperimentResults}. We see that the median values are close to the true values, but there is deviation between the two in Questions Q6--10, which happen to be dot-counting questions. In any case, the log-normal fit can only be approximate, since all guesses are required to be at least 1 and in case of Q5-10 are integers. None the less, we find the log-normal approximation useful for later calculations. 
We also plot how well the data fits a log-normal distribution in Fig~\ref{fig:QQplots}. We notice that, for most questions, the fit is reasonable or even very good, especially for questions Q6--10. An exception is Q5, where the data is highly peaked around the correct answer, 47. This is because people have the ability to count the correct answer for this question, while other answers are much more difficult to infer. That being said, Fig.~\ref{fig:CDFGuess} shows us that the normalized answers are very similar to each other, thus despite the disagreement, results are still qualitatively similar to the log-normal distribution.

To see if the discrepancy between the guess values and true values affected our model results, we centered answers by $\langle\text{ln}(X)\rangle$ as well as by the true values, as shown in Table~\ref{tab:modelrobust}. We see the main text results (in bold) are very similar regardless of how data is centered. Because we see a difference between guesses for dot questions (Q6--10) and ratio question (Q1--4), we separately fit these data subsets to the model. While some of the parameters differed, the qualitative results remain consistent: the model with $p$ and $r$ fit better than simpler models, and $p$ increases in the social influence condition.

\begin{table*}
\footnotesize{
\caption{Robustness of Model Fits\label{tab:modelrobust}}
\begin{tabular}{|l|c|c|c|c|c||c|c|c|c|}\hline
Data&Centering&Pr(Matches Data)& Pr($LR_R$) & Pr($LR_P$) & Pr($LR_0$) &$\langle r\rangle$ & $\sigma_r$ & $\langle p\rangle$ & $\sigma_p$\\
\hline\hline
\bf Control All Q &\bf   Mean&\bf  0.397 &\bf   0.002 & \bf  $\mathbf{<10^{-5}}$ & $\mathbf{<10^{-5}}$ &\bf   0.28 & \bf  0.02 &\bf   0.05 &\bf   0.02 \\
Control Q$>5$ & Mean&0.547 & 0.023 & $<10^{-5}$ & $<10^{-5}$ & 0.54 & 0.05 & 0.10 & 0.04 \\
Control Q$<5$ & Mean&0.61 & 0.061 & $<10^{-5}$ & $<10^{-5}$ & 0.19 & 0.02 & 0.04 & 0.02 \\
Control All Q & True&0.546 & 0.001 & $<10^{-5}$ & $<10^{-5}$ & 0.33 & 0.02 & 0.06 & 0.02 \\
Control Q$>5$ & True&0.546 & 0.020 & $<10^{-5}$ & $<10^{-5}$ & 0.69 & 0.04 & 0.16 & 0.06 \\
Control Q$<5$ & True&0.556 & 0.036 & $<10^{-5}$ & $<10^{-5}$ & 0.12 & 0.02 & 0.03 & 0.02 \\\hline
\bf  Soc. Inf. All Q & \bf  Mean&\bf  0.472 &$\mathbf{<10^{-5}}$ & $\mathbf{<10^{-5}}$ &  $\mathbf{<10^{-5}}$ &\bf   0.08 & \bf  0.02 & \bf  0.21 & \bf  0.01 \\
Soc. Inf. Q$>5$ & Mean&0.498 & $<10^{-5}$ & $<10^{-5}$ &  $<10^{-5}$ & 0.29 & 0.06 & 0.36 & 0.04 \\
Soc. Inf. Q$<5$ & Mean&0.52 & $<10^{-5}$ & $<10^{-5}$ &  $<10^{-5}$  & 0.06 & 0.02 & 0.15 & 0.02 \\
Soc. Inf. All Q & True&0.504 & $<10^{-5}$ & $<10^{-5}$ &  $<10^{-5}$  & 0.11 & 0.02 & 0.21 & 0.01 \\
Soc. Inf. Q $>5$ & True&0.53 & $<10^{-5}$ & $<10^{-5}$ &  $<10^{-5}$  & 0.35 & 0.05 & 0.39 & 0.04 \\
Soc. Inf. Q $<5$ & True&0.561 & $<10^{-5}$ & $<10^{-5}$ &  $<10^{-5}$  & 0.06 & 0.02& 0.14 & 0.02 \\\hline
\end{tabular}
}
\begin{flushleft}
{\bf Bold} results are in the main text. $\langle.\rangle$ represents average and $\sigma$ represents the standard error.
\end{flushleft}
\end{table*}

\subsection{Statistical Methods}

\subsubsection{Fitting the Decision Model}
We fit the decision model defined in the Findings section using maximum likelihood estimation (MLE). This method, however, only provides a point estimate. In order to determine the error of the model parameters, we bootstrap the data (i.e., sample with replacement $n$ times for data of size $n$) and calculate the MLE values for each parameter. 
We repeat this step $1000$ times to create a parameter distribution, and calculate the standard deviation of this distribution to find parameter error bars.

\subsubsection{Comparing Models}
The decision model we fit to data has two parameters, however simpler decision models may fit the data equally well. To check whether this is true, we compare the log-likelihood of the two-parameter decision model, $\ell_{\text{DM}}$, to a simpler model $\ell_{0}$. Let $n$ be the number of observations, then by Wilks' theorem \cite{Wilks1938}, as $n\rightarrow \infty$, $\ell_{\text{DM}} - \ell_{0}$ should follow a $\chi^2(\ell_{\text{DM}} - \ell_{0},k)$ distribution if the data better matches the simpler model, where $k$ is the difference in the degrees of freedom. In our case, the simpler models have one to two fewer degrees of freedom.

\subsubsection{Agreement with Data}
In order to find out whether our model fits the data well, we compare our model's MLE log-likelihood value to the log-likelihood of data bootstrapped from the model with the same answers ($A_1$ and $A_2$) and parameter values ($p$, and $r$) as the empirical data. We then say the data agrees with the model if the probability the bootstrapped data fits the model worse than empirical data is greater than 0.1. In practice, we find the probability typically exceeds 0.4, thus the model is consistent with the data. To check the robustness of this agreement method, we also took the MLE of the bootstrapped data by refitting it to the model. Results are virtually identical.

\subsection{Calculating Probability Error}
To calculate error bars in probabilities, we assume a uniform prior, and use the Beta distribution to create a posterior probability distribution:
\begin{equation}
    \text{Pr}(\rho)=\frac{\Gamma(S+F+2)}{\Gamma(S+1)\Gamma(F+1)} \rho^{S}(1-\rho)^{F}
\end{equation}
where $S$ are the number of successes, $F$ are the number of failures, and $\rho$ is the estimated probability of successes. This allows us to calculate error bars for $\rho$ even when $S=0$ or $F=0$. In all plots, the point estimation is the MLE: $\hat{\rho}=S/(S+F)$.

\subsection{MLE Equation}

Let's assume we can independently measure the position bias parameter, $p$, and random guess parameter, $r$ (e.g. by using results from the present experiment). Let $n_t$ ($n_b$) and $N_t$ ($N_b$) be the number of times an answer was chosen when it was ranked first (last) and the total number of votes to all answers when the answer was ranked first (last). Also, recall that if $s(A_{\text{best}},A_{\text{worst}})>0.5$, $A_{\text{best}}$ is the better answer. 
To estimate $s(A_{\text{best}},A_{\text{worst}})$, we first define 
\begin{dmath}
    \text{Pr}(A_{\text{best}}~First|A_{\text{best}},A_{\text{worst}},p,r) = r/2+(1-r)(p+(1-p) s(A_{\text{best}},A_{\text{worst}})),
\end{dmath}
\begin{dmath}
    \text{Pr}(A_{\text{best}}~Last|A_{\text{best}},A_{\text{worst}},p,r) = r/2+(1-r)((1-p)(1-s(A_{\text{best}},A_{\text{worst}}))),
\end{dmath}
\begin{dmath}
    \text{Pr}(A_{\text{worst}}~First|A_{\text{best}},A_{\text{worst}},p,r) = r/2+(1-r)((1-p)s(A_{\text{best}},A_{\text{worst}})),
\end{dmath}
and
\begin{dmath}
    \text{Pr}(A_{\text{worst}}~Last|A_{\text{best}},A_{\text{worst}},p,r) = r/2+(1-r)(p+(1-p)(1-s(A_{\text{best}},A_{\text{worst}}))),
\end{dmath}
where $A_{\text{best}}$ and $A_{\text{worst}}$ are the respective answers, and ``$A_x~First(Last)$'' means that answer is ordered first (last). 
The variable $s(A_{\text{best}},A_{\text{worst}})$ is the only unknown. 
Surprisingly, we can infer $s(A_{\text{best}},A_{\text{worst}})$ 
without having to normalize answers, let alone know the answer distribution.
The likelihood function of the model is:
\begin{dmath}
    L(n_t,N_t,n_b,N_b,s,p,r) =\text{Pr}(A_1~First|s,p,r)^{n_t}\ \text{Pr}(A_1~Last|s,p,r)^{N_t-n_t} \text{Pr}(A_2~First|s,p,r)^{n_b}  \text{Pr}(A_2~Last|s,p,r)^{N_b-n_b}
\end{dmath}
The log-likelihood function is therefore
\begin{dmath}
    \ell(n_t,N_t,n_b,N_b,s,p,r) =n_t \text{ln}(\text{Pr}(A_1~First|s,p,r))+ (N_t-n_t)\text{ln}(\text{Pr}(A_1~Last|s,p,r))\\
     + n_b \text{ln}(\text{Pr}(A_2~First|s,p,r))
     + (N_b-n_b)\text{ln}(\text{Pr}(A_2~Last|s,p,r))
\end{dmath}
To find the MLE of $s$, we find the solution to  
\begin{equation}
\frac{\partial}{\partial s}\ell(n_t,N_t,n_b,N_b,s,p,r)=0
\end{equation}
and then solve for $s$. 
The MLE value of $s(A_{\text{best}},A_{\text{worst}})$ can easily be solved numerically. 
As long as a researcher records $n_t$, $n_b$, $N_t$, and $N_b$, we can accurately infer answer quality. 

\subsection{Simulation Robustness}

\begin{figure*}[tbh!]
\centering
\includegraphics[width=0.95\textwidth]{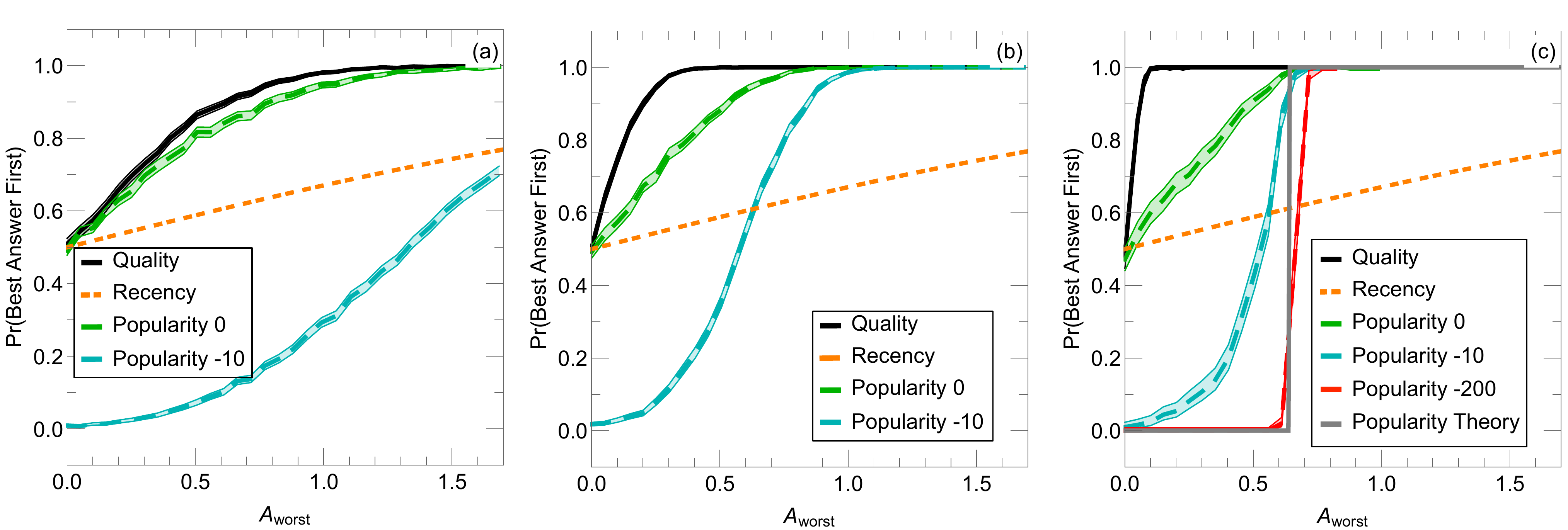}
\caption{Comparison of ranking policies when the quality ranking method incorrectly assumes $r=0.0$ when $r=0.09$. Plots show the probability the best answer is in first place after (a) 50, (b) 500, and (c) 20,000 votes, when answers are ranked by quality, recency, and popularity. ``Popularity 0'', ``Popularity -10'', and ``Popularity -200'' means that the worst answer starts with a 0, 10, or 200 vote advantage, respectively. Quality ranking (dashed lines) consistently ranks the best answer first, even when the two answers have similar quality ($A_{\text{worst}}$<0.6). Recency tends to perform worst, while popularity is sensitive to the initial number of votes answers start with. Even when both answers start off with equal votes, however, popularity-based ranking does not guarantee that the best answer rises to the top. In comparison, quality-based ranking improves with the number of votes an answer accumulates.
\label{fig:rzero} 
}
\end{figure*}

Simulations in the main text are for the case where the quality ranking algorithm correctly assumes $p=0.2$ and $r=0.09$. We explore what happens if one or both assumptions are wrong. For example, we show in Fig.~\ref{fig:rzero} the case when quality ranking assumes $r=0.0$ when $r=0.09$. 
Comparing to Fig.~\ref{fig:InitialGuessModelRankData}, we see results are quantitatively very similar. This is intuitive as randomly choosing the first or last answer with equal probability should not substantially affect their relative ranking. 

\begin{figure}[tbh!]
\centering
  \includegraphics[width=1\textwidth]{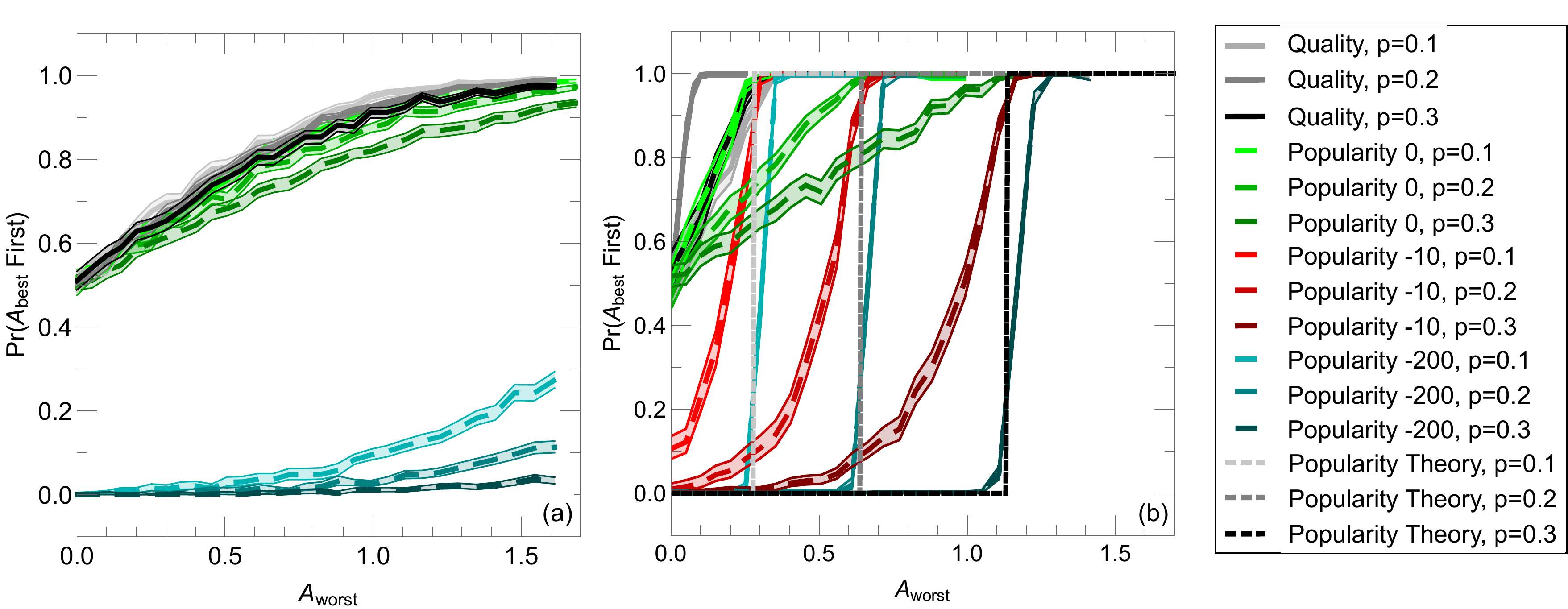}
\caption{\label{fig:SimPRobust} Ordering answers by quality and popularity after (a) 20 and (b) 20K votes.  ``Popularity 0'', ``Popularity -10'', and ``Popularity -200'' means that the worst answer starts with a 0, 10, or 200 vote advantage, respectively. We show different cases when $p=0.1,~0.2,$ or $0.3$ and $r=0.09$. In all cases, quality ranking assumes $p=0.2$ and $r=0.0$, as a worst-case scenario. 
Consistent with the main text, quality ranking (dashed lines) consistently ranks the best answer first, and typically outperforms popularity ranking. When $p=0.1$ and answers begin with equal votes, quality ranking and popularity ranking are comparable. 
}
\end{figure}

What about if we incorrectly estimate both $r$ and $p$? For example, we assume $r=0$ and $p=0.2$, but in actuality, $r=0.09$ and $p=0.1$, $0.2$, or $0.3$? As shown in Fig.~\ref{tab:modelrobust}, the correct $p$ value could vary between $0.1$ and $0.3$ for the social influence condition. 
Results are shown in Fig.~\ref{fig:SimPRobust}. 
Overall, we find that quality ranking still significantly outperforms popularity-based ranking. The only exception is when answers begin with equal votes and $p=0.1$. In this case, quality ranking is comparable after 20 votes, and slightly worse after 20K votes. A website designer could create small-scale experiments to better infer $p$, and after applying a corrected $p$ estimate, they should expect quality ranking to again substantially outperform popularity-based ranking. Overall, even when the quality-ranking algorithm is not correctly parameterized, it still performs rather well, and does not not seem to be very sensitive to the estimate of $p$ or $r$.

\received{January 2020}
\received[revised]{June 2020}
\received[accepted]{July 2020}


\end{document}